\documentclass[twocolumn]{aastex62}
\hypersetup{linkcolor=red,citecolor=blue,filecolor=cyan,urlcolor=blue}
\usepackage{graphicx,color}
\usepackage{amssymb}
\usepackage{amsmath}

\usepackage{natbib}
\usepackage{txfonts}
\usepackage{multirow}
\usepackage{array}
\usepackage{rotating}
\usepackage{sidecap}
\usepackage{hyperref}
\usepackage{epstopdf}
\usepackage{footnote}
\usepackage{tabularx}
\usepackage{booktabs}
\usepackage{longtable}
\usepackage{tabu}
\usepackage{longtable}

\newcommand{\Halpha}{H$\alpha$}

\newcommand{\FeX}{\ion{Fe}{10}}
\newcommand{\FeIX}{\ion{Fe}{9}}

\newcommand{\kms}{km~s$^{-1}$}

\newcommand{\sdo}{\textit{SDO}}
\newcommand{\iris}{\textit{IRIS}}

\shortauthors{Panesar et al.}

\shorttitle{Solar Jetlets}
\shortauthors{Panesar et al.}
\begin{document}

\title{Hi-C 2.1 Observations of Jetlet-like Events at Edges of Solar Magnetic Network Lanes}

\correspondingauthor{Navdeep K. Panesar}
\email{panesar@lmsal.com}
%\email{alphonse.sterling@nasa.gov}

\author[0000-0001-7620-362X]{Navdeep K. Panesar}
\altaffiliation{Also, Visiting Scholar at W. W. Hansen Experimental Physics Laboratory,\\ Stanford University, Stanford, CA 94305, USA}

\affil{Lockheed Martin Solar and Astrophysics Laboratory, 3251 Hanover Street, Bldg. 252, Palo Alto, CA 94304, USA}
\affil{Bay Area Environmental Research Institute, NASA Research Park, Moffett Field, CA 94035, USA}
\affil{NASA Marshall Space Flight Center, Huntsville, AL 35812, USA}

\author[0000-0003-1281-897X]{Alphonse C. Sterling}
\affiliation{NASA Marshall Space Flight Center, Huntsville, AL 35812, USA}

\author[0000-0002-5691-6152]{Ronald L. Moore}
\affiliation{Center for Space Plasma and Aeronomic Research (CSPAR), UAH, Huntsville, AL 35805, USA}
\affiliation{NASA Marshall Space Flight Center, Huntsville, AL 35812, USA}

\author[0000-0002-5608-531X]{Amy R. Winebarger}
\affiliation{NASA Marshall Space Flight Center, Huntsville, AL 35812, USA}

	\author[0000-0001-7817-2978]{Sanjiv K. Tiwari}
\affil{Lockheed Martin Solar and Astrophysics Laboratory, 3251 Hanover Street, Bldg. 252, Palo Alto, CA 94304, USA}
\affil{Bay Area Environmental Research Institute, NASA Research Park, Moffett Field, CA 94035, USA}

\author{Sabrina L. Savage}
\affiliation{NASA Marshall Space Flight Center, Huntsville, AL 35812, USA}

\author{Leon E. Golub}
	\affil{Harvard-Smithsonian Center for Astrophysics, 60 Garden St., Cambridge, MA 02138, USA}
	
	\author{Laurel A. Rachmeler}
	\affiliation{NASA Marshall Space Flight Center, Huntsville, AL 35812, USA}
	
	\author{Ken Kobayashi}
	\affiliation{NASA Marshall Space Flight Center, Huntsville, AL 35812, USA}

	\author{David H. Brooks}
\affil{College of Science, George Mason University, 4400 University Drive, Fairfax, VA 22030, USA}
%\affil{Current address: Hinode Team, ISAS/JAXA, 3-1-1 Yoshinodai, Chuo-ku, Sagamihara, Kanagawa 252-5210, Japan}

	\author{Jonathan W. Cirtain}
	\affil{BWX Technologies, Inc., 800 Main St \#400, Lynchburg, VA 24504}
	\author{Bart De Pontieu}
\affil{Lockheed Martin Solar and Astrophysics Laboratory, 3251 Hanover Street, Bldg. 252, Palo Alto, CA 94304, USA}
\affiliation{Rosseland Centre for Solar Physics, University of Oslo, P.O. Box 1029 Blindern, NO–0315 Oslo, Norway}
\affiliation{Institute of Theoretical Astrophysics, University of Oslo, P.O. Box 1029 Blindern, NO–0315 Oslo, Norway}

%	\author{Scott W. McIntosh}
%\affil{High Altitude Observatory, National Center for Atmospheric Research, P.O. Box 3000, Boulder, CO 80307, USA}
	\author{David E. McKenzie}
\affiliation{NASA Marshall Space Flight Center, Huntsville, AL 35812, USA}
	\author{Richard J. Morton}
\affil{Mathematics, Physics \& Electrical Engineering, Northumbria University, Newcastle Upon Tyne, NE1 8ST, UK}
	\author[0000-0001-9921-0937]{Hardi Peter}
\affil{Max Planck Institute for Solar System Research, Justus-von-Liebig-Weg 3, D-37077, G{\"o}ttingen, Germany}
	\author{Paola Testa}
\affil{Smithsonian Astrophysical Observatory, 60 Garden Street, MS 58, Cambridge, MA 02138, USA}

	\author{Robert W. Walsh}
\affil{Jeremiah Horrocks Institute, University of Central Lancashire, Preston, PR1 2HE, UK}
	\author{Harry P. Warren}
	\affil{Space Science Division, Naval Research Laboratory, Washington, DC 20375, USA}

\begin{abstract}

We present high-resolution, high-cadence observations of six, fine-scale, on-disk jet-like events observed by the High-resolution Coronal Imager 2.1 (Hi-C 2.1) during its sounding-rocket flight. We combine the Hi-C 2.1 images with images from  \sdo/AIA, and \iris, and investigate each event's magnetic setting with co-aligned line-of-sight magnetograms from \sdo/HMI. We find that: (i) all six events are jetlet-like (having apparent properties of \textit{jetlets}), (ii) all six are rooted at edges of magnetic network lanes, (iii)  four of the jetlet-like events stem from sites of flux cancelation between majority-polarity network flux and merging minority-polarity flux, and (iv)  four of the jetlet-like events show brightenings at their bases reminiscent of the base brightenings in coronal jets. The average spire length of the six jetlet-like events (9,000$\pm$3000km) is three times shorter than that for \iris\ jetlets (27,000$\pm$8000km). While not ruling out other generation mechanisms, the observations suggest that at least four of these events may be miniature versions of both larger-scale coronal jets that are driven by minifilament eruptions and still-larger-scale solar eruptions that are driven by filament eruptions.  Therefore, we propose that our Hi-C events are driven by the eruption of a tiny sheared-field flux rope, and that the flux-rope field is built and triggered to erupt by flux cancelation.

\end{abstract}

\keywords{Sun: activity --- Sun: chromosphere---  Sun: corona --- Sun: magnetic fields }

\section{Introduction} \label{sec:intro}

Solar jet-like features are narrow upward streams of plasma ubiquitously observed in the solar atmosphere \citep{shibata92,sterling00b,innes11,raouafi16}. Coronal jets appear in coronal holes, active regions and quiet regions \citep{shibata92,nistico10,pucci13,panesar16a,sterling16} and are often observed in extreme ultraviolet (EUV) and X-ray images \citep{shimojo96,alexander99,cirtain07,savcheva07,huang12,moore18}.

 A long-standing widely-held idea for the production of jets of all sizes is that the jet outflow is magnetically driven by a burst of  reconnection  of emerging closed magnetic field with far-reaching ambient magnetic field, with the burst of reconnection occurring suddenly at the current sheet between them when the current sheet has been sufficiently built up by the emergence of the closed field  \citep[e.g.][]{yokoyama95,shibata11}.  In contrast to this picture however, 
high-resolution and high-cadence observations from the \textit{Solar Dynamics Observatory} (\sdo)/Atmospheric Imaging Assembly (AIA; \citealt{lem12}) show that  coronal jets are often driven by the eruption of a \textit{minifilament} (\citealt{sterling15}; also see \citealt{hong11,shen12,adams14,young14a}).  The eruption drives the jet outflow via reconnection with the far-reaching ambient field and, via internal reconnection of the legs of the minifilament-carrying erupting field, produces a jet bright point (JBP) centered on the neutral line where the minifilament was rooted prior to the eruption. Using \sdo/HMI magnetograms, \cite{panesar16b,panesar17,panesar18a} and \cite{mcglasson19} found that flux cancelation usually  builds the sheared/twisted magnetic field in and around the pre-jet minifilament and triggers it to erupt. 

In addition to coronal jets, smaller-scale network jets (similar in form to coronal jets but 3-4 times smaller), named  \textit{jetlets} by  \cite{raouafi14}, occur at the edges of lanes of the magnetic network  \citep{panesar18b}. %were observed at the base of plumes (see also \citealt{avallone18}).}
They have been seen to originate at canceling neutral lines and to show base brightenings during jet onset \citep{panesar18b}. \cite{panesar18b} found that jetlets are about three times smaller in base width ($<$5,000 km) than typical coronal jets ($\sim$18,000 km).  Therefore, jetlets are plausibly small-scale versions of both larger coronal jets and the still-larger CME-producing eruptions \citep{sterling18}. However, we cannot rule out that there might be some jets of various sizes that are not driven by the eruption of a flux-cancelation-built-and-triggered minifilament flux rope.  Instead, some might be driven by emerging closed magnetic field via reconnection with ambient far-reaching field, as has long been proposed and modeled \citep{shibata92,yokoyama95,shimojo00,shibata11}.

Here, we present even-higher-resolution EUV observations of still-smaller jet-like events observed by NASA's High-resolution Coronal Imager 2.1 (Hi-C 2.1; hereafter `Hi-C') on a sounding rocket. Hi-C's images reveal fine-scale evolving structures that have not been discerned before at Hi-C's observing wavelength. During the five minutes of observations of Hi-C, we identified six jet-like fine-scale events (Table \ref{tab:list}). To judge whether they are still-smaller versions of \iris\ jetlets and coronal jets, we investigate the magnetic setting of these six jet-like events, examine their physical properties using the Hi-C data, and compare their properties with those of \iris\ jetlets.

\floattable
%	\begin{center}

\begin{table}
	
	\setlength{\tabcolsep}{1.0pt} %% to adjust space between columns
	
	%	\tabletypesize{\scriptsize}
	\scriptsize{
		\caption{Measured parameters of observed Hi-C  jet-like events\label{tab:list}}
		% \tablewidth{100pt}
		\renewcommand{\arraystretch}{1.0}% for Tighter rows
		\begin{tabular}{c*{7}{c}}
			\noalign{\smallskip}\tableline\tableline \noalign{\smallskip}
			%\multicolumn{10}{l}
		%%%%%%%%%%%%%%%%%%%%%%%%%%%%%%%%%%%%%%%%%%%%%%%%%%%%%%%%%%%%%%%%%%%	
			Event\tablenotemark{a} &   Type\tablenotemark{b}   & Spire Length\tablenotemark{c} & Spire Width\tablenotemark{d}   & Speed\tablenotemark{e} & Base\tablenotemark{f} & Discernible\tablenotemark{g}   \\
		%%%%%%%%%%%%%%%%%%%%%%%%%%%%%%%%%%%%%%%%%%%%%%%%%%%%%%%%%%%%%%%%%%%		
			No.   &    &  (km) &  (km) &  (\kms) &  Brightening & Minority flux  \\
			
			\noalign{\smallskip}\hline \noalign{\smallskip}
			
		%%%%%%%%%%%%%%%%%%%%%%%%%%%%%%%%%%%%%%%%%%%%%%%%%%%%%%%%%%%%%%%%%%%	
		1  & jetlet-like & 12000$\pm$800 &  750$\pm$50  &110$\pm$30  & No & Yes \\ 
			
		%%%%%%%%%%%%%%%%%%%%%%%%%%%%%%%%%%%%%%%%%%%%%%%%%%%%%%%%%%%%%%%%%%%	
			
		2  & jetlet-like & 14000$\pm$300  & 600$\pm$100 & 24$\pm$3  & No &  Yes \\ 
			
		%%%%%%%%%%%%%%%%%%%%%%%%%%%%%%%%%%%%%%%%%%%%%%%%%%%%%%%%%%%%%%%%%%%
			
		3\tablenotemark{h}  & jetlet-like & 9000$\pm$1000 & 750$\pm$100  & 110$\pm$20 & Yes & Yes\\ 
		
	    %%%%%%%%%%%%%%%%%%%%%%%%%%%%%%%%%%%%%%%%%%%%%%%%%%%%%%%%%%%%%%%%%%%	
			
		4   &jetlet-like & 10000$\pm$650  &  650$\pm$50 &60$\pm$10 & Yes &  Yes \\
			 
	    %%%%%%%%%%%%%%%%%%%%%%%%%%%%%%%%%%%%%%%%%%%%%%%%%%%%%%%%%%%%%%%%%%%	
			
		5   & jetlet-like & 5000$\pm$1000 &  400$\pm$50 &15$\pm$5  & Yes & No \\ 
			
		%%%%%%%%%%%%%%%%%%%%%%%%%%%%%%%%%%%%%%%%%%%%%%%%%%%%%%%%%%%%%%%%%%%	
	
		6   & jetlet-like & 6000$\pm$1000 &  350$\pm$50 & 50$\pm$10 & Yes & No \\ 
		
		%%%%%%%%%%%%%%%%%%%%%%%%%%%%%%%%%%%%%%%%%%%%%%%%%%%%%%%%%%%%%%%%%%%

			\noalign{\smallskip}\tableline\tableline \noalign{\smallskip}
		average$\pm$1$\sigma$$_{ave}$ & & 9000$\pm$3000 & 600$\pm$150 & 60$\pm$40\\
	\hline
			
		\end{tabular}
		
		\tablenotetext{a}{Locations of observed events are shown in Figure \ref{fig0}.}
		\tablenotetext{b}{Type based on their physical properties; see text.} 
		\tablenotetext{c}{Maximum length of the spire measured in Hi-C images from the base to the visible tip near time of maximum extent. }
		\tablenotetext{d}{Width measured in the middle of the spire near time of maximum extent in Hi-C images.}
		
		\tablenotetext{e}{Plane-of-sky speed along the  spire. Speeds and uncertainties were measured from  Hi-C 172\AA\ time-distance maps.}
			\tablenotetext{f}{Whether base brightening is discernible in Hi-C 172\AA\ and AIA 171\AA\  images. Base brightenings in jetlet-like events(3,4) are clearly visible than the base brightenings in jetlet-like events(5,6).}
				\tablenotetext{g}{Whether minority-polarity flux is discernible at the base of the jetlet-like events.}
		\tablenotetext{h}{This jetlet-like event starts before the Hi-C coverage. All measurements were done using AIA 171\AA\ images. }
		
	}
	
\end{table}

	%\end{center}

\begin{figure*}
	\centering
	\includegraphics[width=\linewidth]{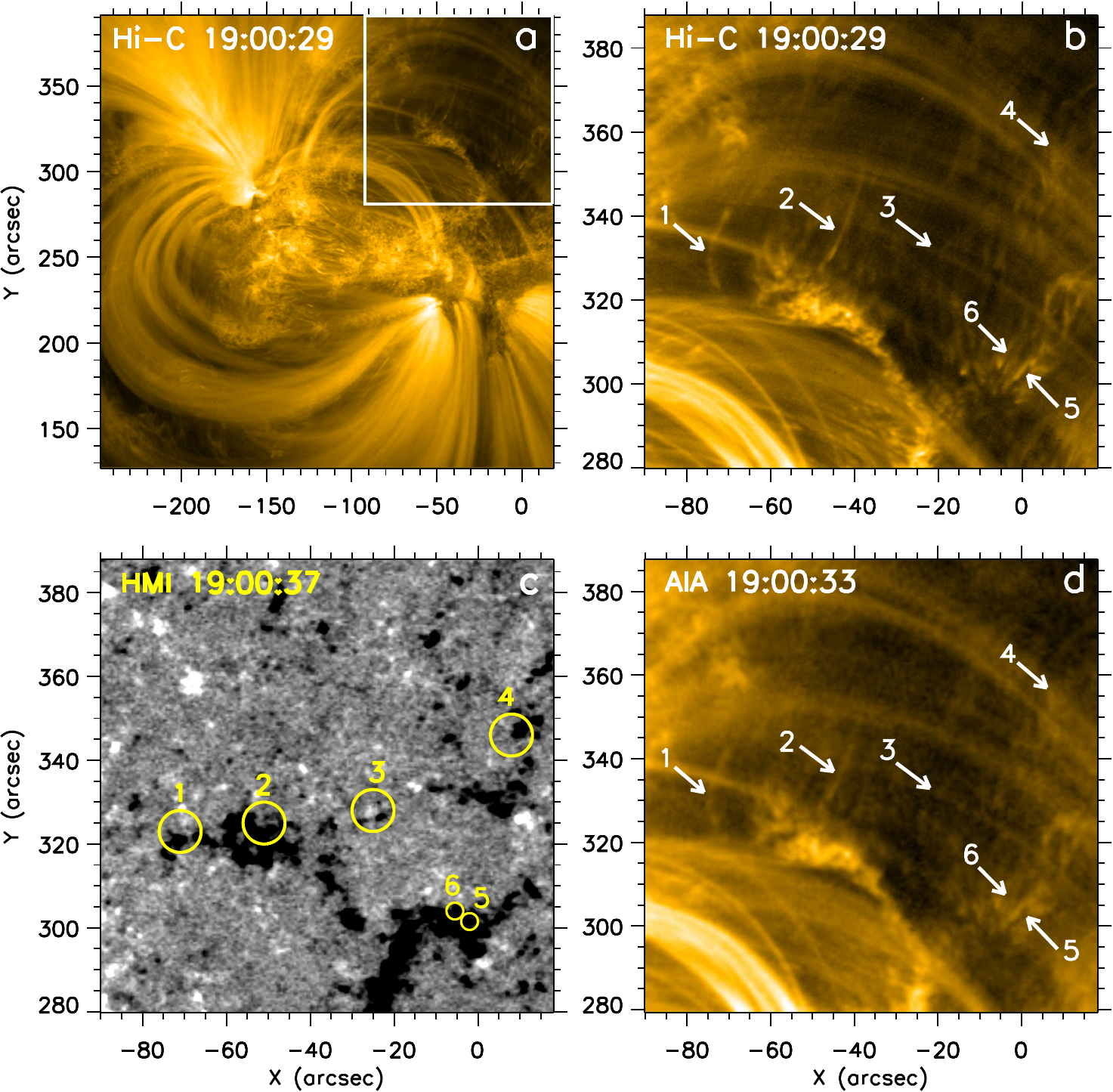}
	\caption{ Locations of the six jetlet-like events of Table \ref{tab:list}: Panel (a) shows Hi-C's full FOV of AR 12712 on 2018-May-29, in 172\AA\ images; the AR was located near solar disk center (N15, E10); the white box in (a) outlines the FOV shown in panels (b), (c), and (d). Panels (b), (c) and (d) show the zoomed-in Hi-C image, an HMI magnetogram, and an AIA 171 \AA\ image, respectively, of that region. The white arrows and labels (number 1--6) in (b) and (d) point to the locations of the six events. The yellow circles in (c) encircle the photospheric magnetic flux at the base of the spire in each event. Jetlet-like events 3 and 4 are not seen in Figure \ref{fig0} due to their different times of appearance.} \label{fig0}
\end{figure*}

\section{DATA SET}\label{data} 

Hi-C was launched on a sounding rocket on 2018 May 29, and observed an active region (AR 12712) in 172\AA\ emission (including the \FeIX/\FeX\ spectral lines) for about 5 min (18:56:26--19:01:43UT). The Hi-C images have a pixel size of 0\arcsec.129 and have 4s cadence \citep{rachmeler19}.  We identified six fine-scale jet-like events (Figure \ref{fig0}a) in the AR's outskirts, in the north-west of the Hi-C field-of-view (FOV). Here we study their structure, evolution, and magnetic settings in detail. 

We have coordinated data from the \textit{Interface Region Imaging Spectrograph} (\iris; \citealt{pontieu14}), and AIA \citep{lem12} and HMI \citep{scherrer12} from \sdo. In AIA images, our six Hi-C events are also seen in 304 and 171\AA;  they are barely visible in 193\AA\ and 211\AA\ and invisible in other AIA channels. Here we present only  AIA 171\AA\ images because they show best the Hi-C 172\AA\ events.

Out of the six jet-like events, only one (event 1 of Table \ref{tab:list})  was observed in \iris\ slit-jaw images (SJIs), because of limited overlap of the \iris\ FOV and the Hi-C FOV. The \iris\ spectral slit did not cover any of these events. We  use only Si IV  SJIs because the jet-like events are best seen in this bandpass. AIA images, Si~IV SJIs, and HMI magnetograms were co-aligned with the Hi-C data. We estimate the co-alignment to be within about $1''$. To enhance the visibility of weak minority-polarity  flux near the network lanes, we summed two magnetograms at each time step, the one taken at that step and the subsequent one taken 45s later. 
%\sdo\ and \iris\ data sets are co-aligned with Hi-C data. \sdo\ data is also corrected to the Hi-C roll-angle and our final \sdo-Hi-C alignment to be accurate to $\lesssim$1" for our events. AIA images, Si~IV SJIs, and HMI magnetograms were co-aligned with respect to the Hi-C data, including an adjustment of the roll angle of \sdo\ and \iris\ to match that of Hi-C. We estimate the co-alignment to be within about 1''. 

\begin{figure*}
	\centering
	\includegraphics[width=\linewidth]{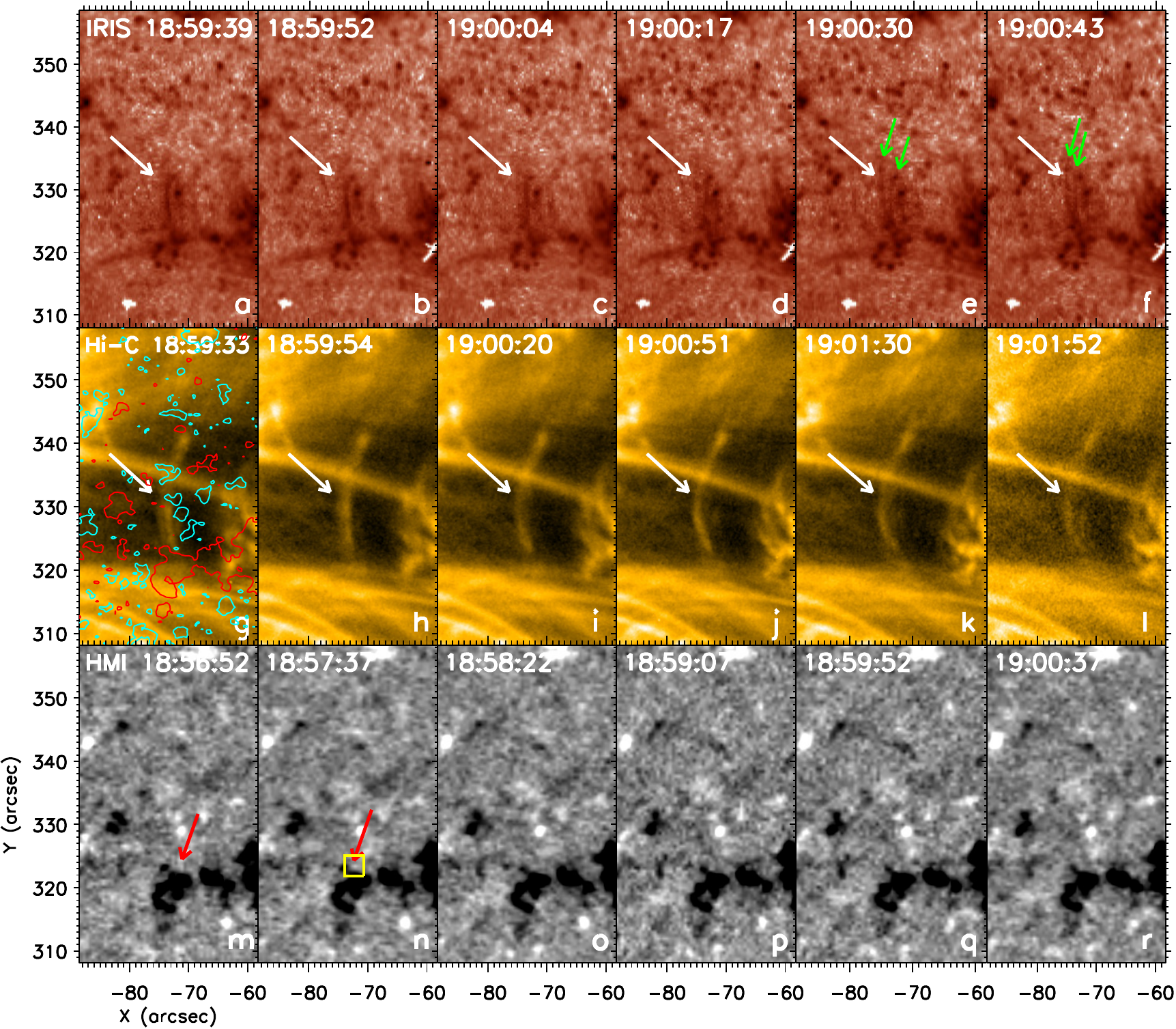}
	\caption{ Hi-C jetlet-like event 1 of Table \ref{tab:list}. Panels (a-f) and (g-l) show the \iris\ Si IV SJIs (in reverse color) and Hi-C 172 \AA\ images of the event, respectively. The white arrows point to the spire.  The green arrows in (e) and (f) point to the two strands of the spire. Panels (m-r) show HMI magnetograms of the same region. The red arrows in (m,n) point to a faint canceling positive-polarity flux grain. The yellow box in (n) shows the region measured for the magnetic flux time plot in Figure \ref{fig4}a. Turquoise and red ($\pm$20-Gauss) contours (at 18:58:22UT) in (g) outline the positive-and negative-polarity flux patches, respectively.  Animations (MOVIE1-iris, MOVIE1-hic and MOVIE1-hmi) of this figure are available.} \label{fig1}
\end{figure*} 
%The details of these jetlets are listed in Table~\ref{tab:list}.

\section{Results}

\subsection{Overview}\label{over}

 Figure \ref{fig0}b shows the Hi-C FOV covering our six jet-like events. %Out of the six, we identified four jetlet-like events (their properties are similar to \iris\ jetlets of \citealt{panesar18b}) and two spicule-like events (they have properties typical of chromospheric spicules \citep{pereira12}; Table \ref{tab:list}). 
 We determined that their properties are similar to \iris\ jetlets of \citealt{panesar18b}. 
 Figure \ref{fig0}c shows that these events occur at the edges of magnetic network lanes. These jetlet-like events occur at the base of far-reaching coronal magnetic loops. All six jetlet-like events  appear in AIA 171\AA\ images, but not as clearly as in Hi-C images (Figure \ref{fig0}d); thus  we likely would not have noticed these features if we observed them only in AIA 171 \AA\	images, without first having examined the higher-resolution Hi-C images.  Out of the six events, here we present four jetlet-like events (Section \ref{res1})  in detail.
 
 %hereafter, ``jetlet"; 
\begin{figure*}
	\centering
	\includegraphics[width=\linewidth]{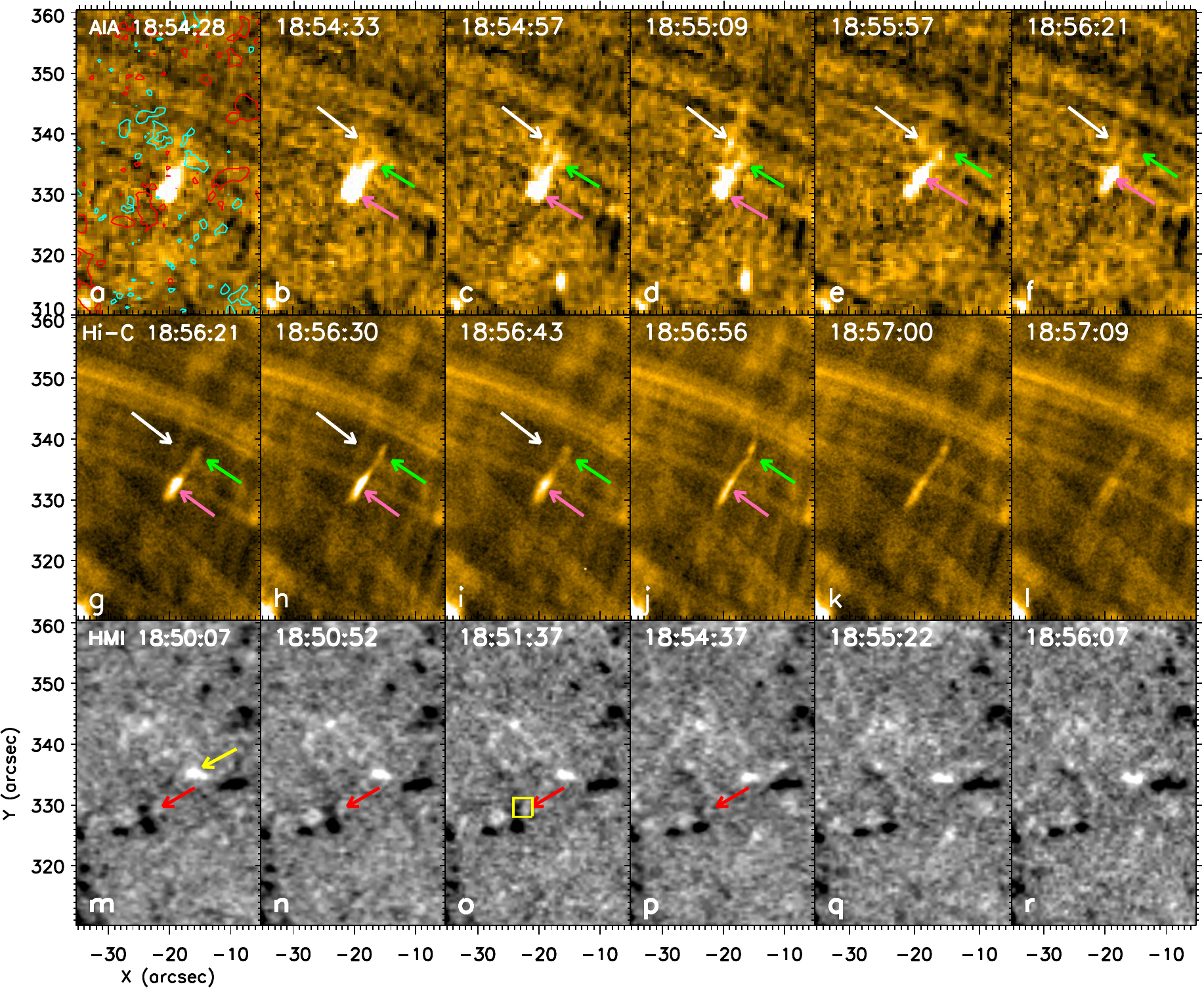}
	\caption{ Hi-C jetlet-like event 3 of Table \ref{tab:list}. Panels (a-f) and (g-l) show AIA 171 \AA\ fixed-difference images and Hi-C images of the event. The white and pink arrows point to the spire and to the base brightening, respectively. The green arrows point to the loop-like brightening that brightens during the eruption. Panels (m-r) show the HMI magnetograms of the same region. The red arrows in (m-p) point to the canceling  minority-polarity flux grain. The yellow arrow in (m) points to the neighboring minority-polarity flux patch at the far end of the larger bright base loop. The yellow box in (o) shows the region measured for the magnetic flux time plot in Figure \ref{fig4}b. Turquoise and red ($\pm$20-Gauss) contours (at 18:54:37UT) in (a) outline positive and negative polarity flux patches, respectively. Animations (MOVIE3-aia, MOVIE3-hic and MOVIE3-hmi) of this figure are available.} \label{fig2}
\end{figure*} 

\subsection{Jetlet-like Events}\label{res1}
\subsubsection{Jetlet-like Event 1}

Figures \ref{fig1}(a-f) and (g-l) show the \iris\ SJIs and Hi-C 172\AA\ images, respectively of  jetlet-like event 1. Plasma starts to move upwards along a pre-existing faint loop at about 18:59:15UT (Movie1-hic), and appears as a bright and thin spire along the loop over 19:00:20 to 19:00:51 UT (Figure \ref{fig1}). At the same time the spire becomes prominent in \iris\ Si SJIs (Movie1-iris and Figure \ref{fig1}a-f). \iris\ SJIs show two separate strands in the spire (green arrows in Figures \ref{fig1}e and f). The decrease in separation of the two strands from 19:00:30 UT to 19:00:43 UT is a possible but not decisive signature of twisting/spinning of the spire. After 19:01:52 UT, the spire starts to disappear and the loop structure persists (Movie1-hic).  We do not see, rising from the base as the event begins, any dark feature that might be taken to be a tiny erupting filament; that is, no erupting `minifilament/microfilament' is discernible in these data. 

\begin{figure*}
	\centering
	\includegraphics[width=\linewidth]{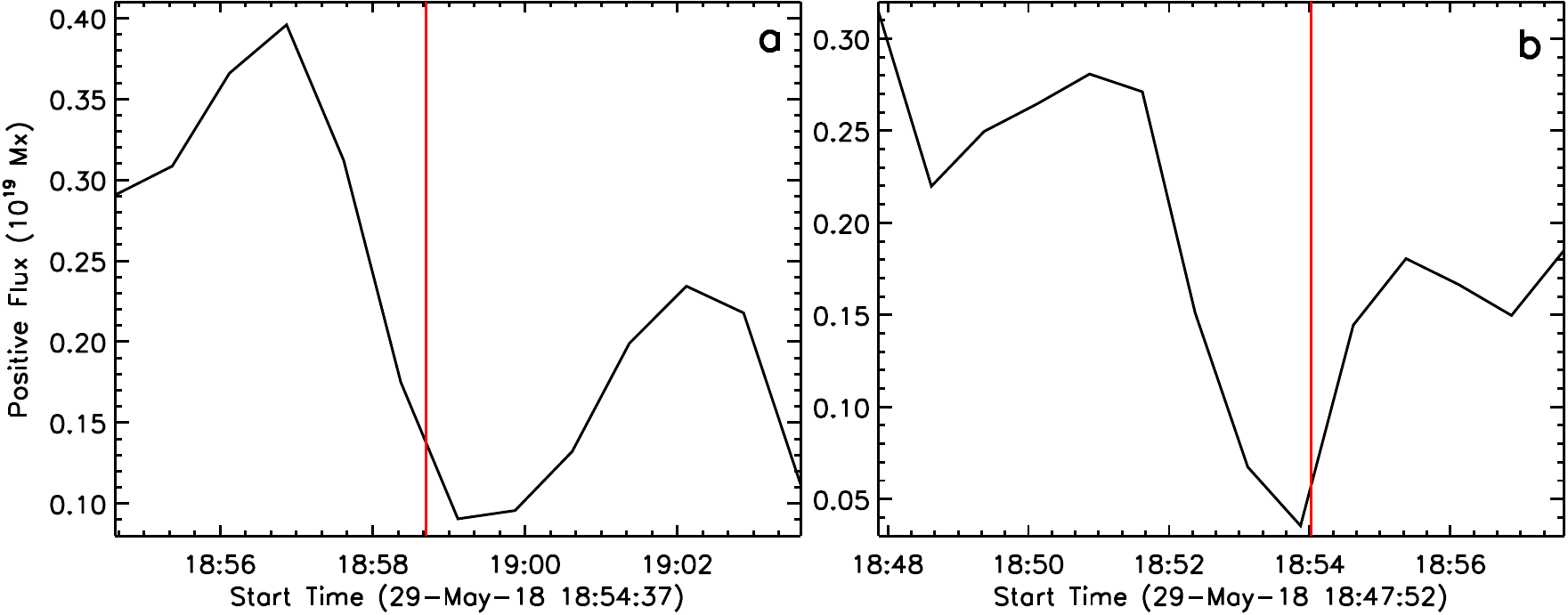}
	\caption{ Magnetic flux plots for jetlet-like event 1 and jetlet-like event 3 of Table \ref{tab:list}. Panels (a) and (b) show the positive flux as function of time computed inside the yellow box region of Figure \ref{fig1}n and Figure \ref{fig2}o, respectively. The red lines mark the event onset times.} \label{fig4}
\end{figure*}

Figure \ref{fig1}(m-r) displays the line-of-sight (LOS) photospheric magnetic flux at the base. The jetlet-like event is rooted at the edge of the negative-polarity network flux lane, between the (majority-polarity) network flux  and a smaller weak minority-polarity (positive) flux patch (red arrows in Figures \ref{fig1}m,n). We followed these fluxes and observed discernible flux cancelation of the minority-polarity flux patch at the neutral line. To examine the magnetic field evolution quantitatively, we measured the minority-polarity flux patch of the base region that is bounded by a yellow box region of Figure \ref{fig1}n. Figure \ref{fig4}a shows a decrease in that positive flux  between 18:57:37 and 19:00:37UT, presumably from flux cancelation that may have triggered the eruption of the jetlet-like event 1, as in larger jetlets and coronal jets examined by \cite{panesar16b,panesar17,panesar18a,panesar18b} and by \cite{mcglasson19}.  This event is similar to \iris\ jetlets \citep{panesar18b} in that the jetlets also occurred at the edges of network flux lanes at canceling neutral lines. Unlike the \cite{panesar18b} jetlets, this jetlet-like event does not show any base brightening (corresponding to the JBP in coronal jets/jetlets) at the canceling neutral line, either in the Hi-C or in the \iris\ images.

\subsubsection{Jetlet-like Event 3}
Figure \ref{fig2} and movies (Movie3-hic, Movie3-aia and Movie-hmi) show the evolution of jetlet-like event 3. Hi-C images do not cover the beginning phase of this jetlet-like event, therefore we supplement Hi-C with AIA 171\AA\ images. There are two jetlet-like events from the same network edge: the first one, which ended prior to the start of Hi-C coverage, starts at 18:52:21UT (not listed in Table \ref{tab:list}), and the  second one starts at 18:54:09UT. Sequential coronal jets have also been observed to erupt multiple times from the same neutral line during ongoing flux cancelation there \citep{sterling17,panesar17}. Here, we present only the second one (jetlet-like event-3), the one partly covered by Hi-C (although both are visible in Movie3-aia). In both of these two jetlet-like events, a faint spire appears next to a small bright loop (green arrows in Figure \ref{fig2}). This small loop brightens as the spire shoots out.
 
For the second of these two jetlet-like events, the base starts to brighten at $\sim$18:54:09UT (Movie3-aia). The brightening sits on the neutral line between a larger majority-polarity (negative) flux clump and smaller and weaker minority-polarity (positive) flux grains (Figures \ref{fig2}a and \ref{fig2}m-p). This base brightening (pink arrows in Figure \ref{fig2}) appears to be a miniature version of the jet-base bright points (JBPs) \citep{sterling15,panesar16b,panesar18b} that  occur at the canceling neutral line at the base of \iris\ jetlets and coronal jets. 
 However, this jetlet-like event does not show  an erupting minifilament.
 Later, starting at 18:54:33UT (Figures \ref{fig2}(b-i)), a faint-spire extends up into the corona with an average speed of 115$\pm$20 \kms. Simultaneously, some additional loop-like brightening appears next to the spire (green arrows in Figure \ref{fig2}) and extends to the neighboring positive-polarity flux patch (Figures \ref{fig2}a and \ref{fig2}m). The spire fades away after 18:56:56UT, in both AIA and \iris\ images. 
 
Figure \ref{fig4}b shows the positive flux plot versus time. The positive flux starts to decrease at $\sim$18:51:37UT and jetlet-like event starts at  $\sim$18:54 UT. One can see in Figures \ref{fig2}(m-r) that the positive (and negative) flux decreases at the base of the jetlet-like event. 

% \textbf{This brightening is similar to the \textit{external brightenings} \citep{sterling15} in coronal jets, brightenings that connect to neighboring opposite-polarity flux patches and are made by external reconnection of the erupting arcade that wraps around the erupting minifilament flux rope.}
 
 %The flux cancelation starts before the jetlet onset at $\sim$18:45:37UT and continues for 10 minutes (until 18:56:07UT, Movie3-hmi), until the minority-polarity (positive) flux patch disappears. The flux cancels at the neutral line in the jetlet-base-region, and the duration of the cancelation is consistent with flux cancelation preparing and triggering each of the two sequential eruptions. Weak flux grains continue to cancel after the first jetlet (at 18:54:37UT, Movie3-hmi), and that further cancelation leads to the final jetlet. Sequential coronal jets have also been observed to erupt multiple times from the same neutral line due to progressive flux cancelation \citep{sterling17,panesar17}.

\subsubsection{Jetlet-like Events 5 and 6}\label{res2}

\begin{figure*}
	\centering
	\includegraphics[width=\linewidth]{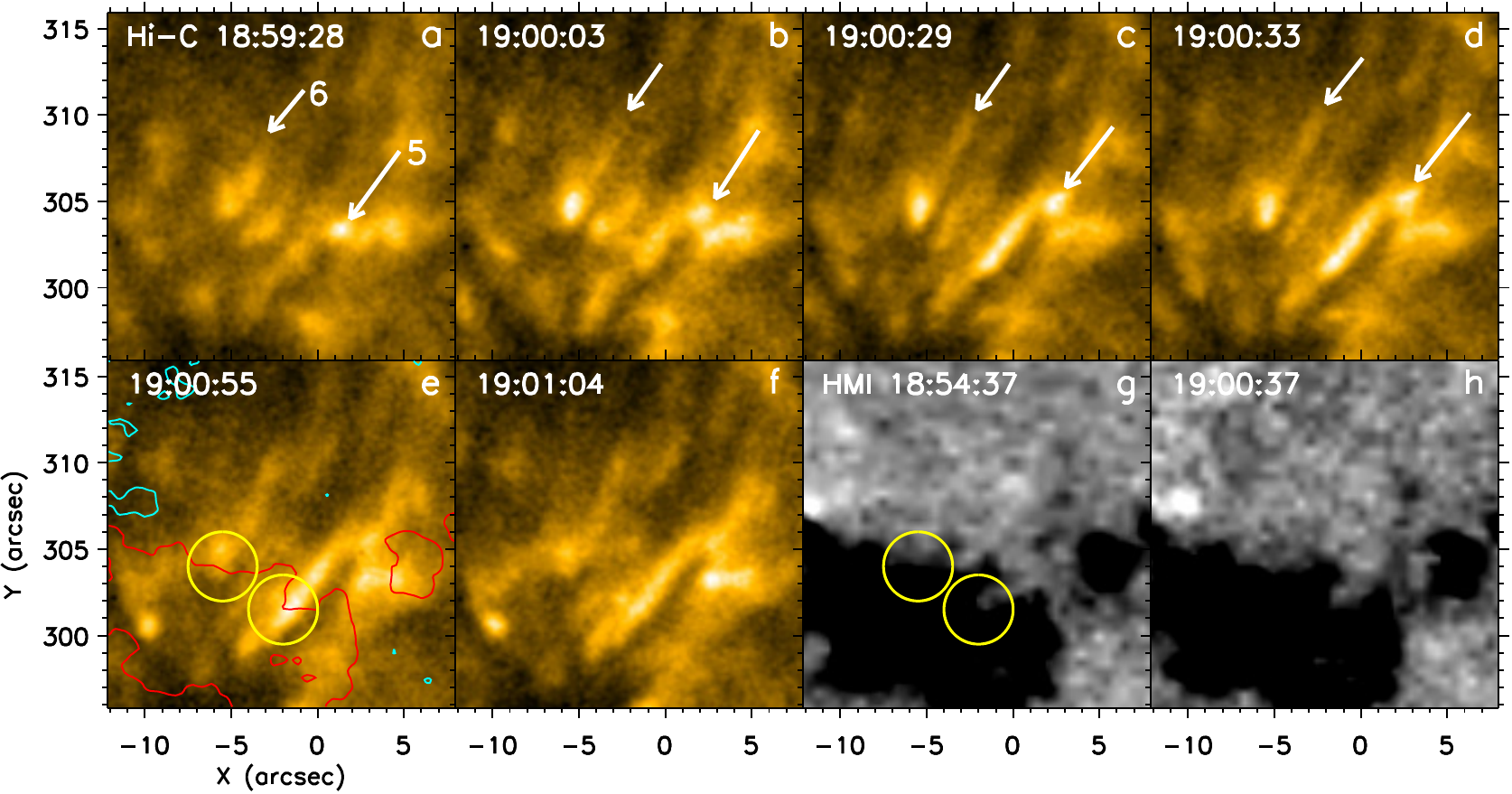}
	\caption{ Hi-C jetlet-like event 5 and jetlet-like event 6 of Table \ref{tab:list}. Panels (a-f) show the Hi-C 172 \AA\ images of the jetlet-like events. The right and left white arrows point to jetlet-like events 5 and jetlet-like event 6, respectively.
		Panels (g and h) show the HMI magnetograms of the same region. Each yellow circle centers on the base of the corresponding event. Turquoise and red ($\pm$20-Gauss) contours (at 19:00:37UT) in (e) outline positive and negative polarity flux patches, respectively. Animations (MOVIE5-hic and MOVIE5-hmi) of this figure are available.} \label{fig3}
\end{figure*}

Events 5 and 6 are different from the other four events. They are shorter in length and narrower in width (Table \ref{tab:list}). The average of the observed widths (375$\pm$75 km) , lengths (5500$\pm$700 km), and speeds (32$\pm$20 km) are similar to spicule widths ($\leq$400 km), lengths (3000-6000 km) and speeds (10-100\kms) of \cite{pontieu07,pereira12} and \cite{tian14}. The spires of these two events  appear as a dark structure adjacent to a bright strand (Figures \ref{fig3}a-d and Movie5-hic) and are rooted at the edge of a clump of network-lane flux (Figures \ref{fig3}g,h and Movie5-hmi). However, we  find no definite signature of minority-polarity flux in these events. 
%, at places where positive-polarity flux grains might be present but too weak to be detected by HMI.

At 18:59:02UT (Movie5-hic), a small brightening appears at the base of jetlet-like event 5. After the start of the brightening the spire rises with a speed of 15$\pm$5 \kms, with a bright tip at its front (Figure \ref{fig3}a-d). In jetlet-like event 6, base  brightening appears at 18:59:37UT (Movie5-hic). The spire  extends out with a  speed of 50$\pm$10 \kms. In both events, a bright component appears, to the left of the dark part of the spire (Figure \ref{fig3}b), and grows along with the dark part.  However, we are not certain whether there is a direct connection between the bright and dark strands. Another possibility is that these dark structures are EUV absorption components of common chromospheric jets, with surrounding brightenings due to hot tips of chromospheric jets, and other transition region brightenings  \citep{skogsrud15,pontieu11}. Moreover, the brightenings of these two events appear different from the other four events, appearing adjacent to the dark features	and lasting for a relatively short time.  Whether this is due to an intrinsic difference among our events, or due to resolution and sensitivity limitations of Hi-C, cannot be determined from this data set alone.

%In both events, the spire appears to consist of two strands: a dark strand and a bright strand along the left side of the dark strand.  The bright strand appears to grow along with the dark strand.  That is, the bright strand appears to be a component of the spire, co-generated with the dark strand.  [An alternative possibility is that the spire is only the dark strand, which is dark because it  is a chromospheric-temperature jet (a chromospheric spicule) that blocks background 172 Å emission  \citep{skogsrud15,pontieu11}, and the bright strand is simply adjacent background 172 Å emission.]  The base brightenings in these two events are different from those in the other four events in that they are offset to one side of the bottom of the spire and are not as long-lasting.

%\\textbf{Conceivably, the base brightening resulted from tiny eruptions \citep{sterling16b} prepared and triggered by flux cancelation but that cancelation is not detected by HMI due to its limited resolution and sensitivity. } 

\section{Discussion}

Using high-resolution ($\sim$150 km), high-cadence (4.5s) Hi-C 2.1 images, we examined the evolution of six small-scale  jetlet-like events  along with their magnetic setting. We find that (i) the Hi-C jetlet-like  events are rooted at edges of magnetic network lanes similar to \iris\ jetlets \citep{panesar18b}; and (ii) jetlet-like events (1-4) stem from sites of flux cancelation between merging majority-polarity and minority-polarity flux patches, evocative of coronal jets; and (iii) jetlet-like events 3, 4, 5 and 6 show brightenings at their bases, reminiscent of the base brightenings in coronal jets. As described in Section \ref{res2}, because of the  variable quality of the data the base brightenings in  jetlet-like events(5,6) are not as clearly visible as in  jetlet-like events(3,4). With the above cautions regarding the base-brightening of jetlet-like events 5 and 6, overall these results are consistent with (although only marginally so for events 5 and 6) our Hi-C events being smaller versions of IRIS jetlets and coronal jets.

  We find evidence both consistent with and not consistent with these events being scaled-down versions of coronal jets and jetlets.  The Hi-C jetlet-like events (1-4), on average, are at least three times smaller in spire length (10,000$\pm$3000 km) and five times smaller in spire width (650$\pm$70 km) than the spire length (27,000$\pm$8000 km) and spire width (3200$\pm$200 km) of \iris\ jetlets \citep{panesar18b}. The average speeds of the Hi-C jetlet-like events (70$\pm$30 \kms) are similar to the speeds of \iris\ jetlets (75$\pm$40 \kms) and of coronal jets  (100$\pm$30 \kms; \citealt{panesar16b,panesar18a}). These measured values are additional evidence that the Hi-C jetlet-like events are scaled-down versions of larger jetlets and coronal jets. 

Jetlet-like events (1-4) occur at sites of discernible flux cancelation. In jetlet-like events 3 and 4, we also observe brightenings at the canceling neutral line. This is consistent with flux cancelation at the neutral line building and triggering these  jetlet-like eruptions (and the jetlet-like eruption preceding jetlet-like event 3). 

The evidence that our Hi-C jetlet-like events are not consistent with coronal jets (and also perhaps jetlets) includes that we do not find any evidence of an erupting minifilament in the Hi-C events. Also, not all of our Hi-C events include brightenings that would correspond to the JBP of coronal jets.  One possibility is that these Hi-C events  are merely too small to resolve possible minifilament counterparts and the brightenings are simply be too weak for Hi-C to detect. We plan to analyze similar small-scale events using \iris\ spectra to find out if there is any cool material in jetlet-like events. 

Alternatively, the Hi-C events might instead be different from coronal jets and jetlets (also see Section \ref{res2}). They might, for example,  be driven by some other mechanism(s), such as (i) evaporation flow in which a jet can shoot out by the gas pressure gradient during the eruption; (ii) magnetic pressure force driving  a jet as a result of magnetic reconnection between a twisted loop and ambient coronal field; (iii) or a jet may result from magnetic tension in the interchange-reconnection outflow when the current sheet for that reconnection is made by field emergence rather than by eruption of the minifilament-carrying field \citep[e.g.][]{shibata96,shibata99, shimojo00}. 
	
	  More studies are required to differentiate between these (and other) possible reasons for the inconsistencies between our Hi-C events on the one hand, and coronal jets and jetlets on the other hand.

The Hi-C 172\AA\ jetlet-like events (5,6) show lengths, widths, and  speeds in the range of chromospheric spicule lengths (3000-6000km), widths (300-400km) and  speeds (15-110\kms) observed by \cite{pontieu07,pereira12,skogsrud15}. Both of these events occur at the edges of the network lanes, as do chromospheric spicules. This suggests that some chromospheric-spicule events might have a transition-region coronal temperature component that is detectable in FeIX/X emission by Hi-C, in agreement with AIA EUV spicule observations presented by \cite{pontieu11}. It is possible that small-scale minority-flux was present in the base of these events, at the edge of the network lane, but was not detectable by HMI. Higher spatial resolution, higher magnetic sensitivity data from future telescopes (e.g. DKIST) will further clarify the magnetic setting of such small-scale  events.  

However, the presence of minority polarity at the base of these types of jets is not unique to the minifilament eruption model for jets. In fact, an alternative promising mechanism for spicules \citep{sykora17} also involves the presence of opposite polarity flux at the footpoint of strong flux concentrations and associated spicules. In their model, a sudden release of magnetic tension by ambipolar diffusion impulsively drives small-scale jet outflows from edges of the magnetic network. 
 
We cannot however be certain that we are observing true jetlet-like events. Another possibility is that we are just seeing selected chromospheric fibrils that show strongly in 171 \AA\, just like some dynamic AR \Halpha\ fibrils show up in EUV  \citep{berger99,pontieu99}. There is strong observational  \citep[e.g.][]{pontieu03} and theoretical  \citep[e.g.][]{hansteen06,heggland11} evidence that dynamic fibrils result from magnetic acoustic waves driven from photospheric motions. Such a mechanism is very different from that suspected of driving coronal jets. It is vital to have high-resolution instruments such as Hi-C in conjunction with H$\alpha$, SST and DKIST to differentiate between these disparate ideas for the small-scale features like our Hi-C jetlets. Detailed studies with Hi-C-like instruments in conjunction with high-quality \Halpha\  observations would be needed to distinguish with confidence between these possibilities.

%\textbf{As I mentioned, this brightening at the “base” of jets 5 and 6 is speculative. I think a sentence should be added that makes clear that it is difficult to identify the base/top of events 5/6 and that the causality of brightenings and the jet evolution are difficult to establish.}

Hi-C jetlet-like events(3, 4, 5, 6) show base brightenings during  onset. However, the base brightenings of jetlet-like events(5, 6) are not as clear as the base brightenings of jetlet-like events(3, 4). Therefore the causality of brightenings and the jet evolution are difficult to establish in these two events. The brightenings appear at the neutral line and/or network-lane edge and are analogs of the jet-base brightenings that are seen to occur at and near the cancelation neutral line during coronal-jet onset. If the Hi-C events are miniature versions of coronal jets then these jet-base-like base brightenings would result from  both \textit{internal reconnection} \citep{sterling15} that occurs within the legs of an erupting magnetic arcade with the minifilament flux rope in its core and \textit{external reconnection} of the erupting magnetic arcade with the encountered far-reaching field. The spire would result from the external reconnection driven by the eruption of the minifilament-carrying field. Several simulations of jet eruptions have shown that the jet spire outflow can be driven by the magnetic pressure in the magnetic twist that is transferred from the twisted closed field to the ambient far-reaching field by the interchange reconnection  \citep{shibata86,shibata99,wyper18}.

%(this reconnection occurs when erupting flux-rope arcade reconnects with the neighboring far-reaching field). %In jetlet-3, we observe brightenings in a loop that connects to a nearby opposite-polarity flux patch, plausibly formed by the external reconnection as we often see in coronal jets \citep{panesar16b}.

Jetlet-like events that occur at the footpoints of far-reaching loops (evidently due to flux cancelation) might contribute to coronal heating of the loops. However, a more detailed analysis is required to establish that eruptions from flux-cancelation at the feet of coronal loops drive heating in the loops \citep[e.g.][]{tiwari14,tiwari17,chitta17}.

Our observations  provide evidence that some Hi-C jetlet-like eruptions are analogs of  larger-scale coronal-jet minifilament eruptions and also still-larger-scale solar eruptions that make CMEs. If this is the case, then based on these four Hi-C jetlet-like  events and previous observations of \iris\ jetlets and coronal jets,  the implication would be that flux cancelation may play a key role in the buildup and triggering of solar eruptions of a wide range of sizes, from as small as jetlets to as large as CME eruptions.

%\newpage
\acknowledgments
We thank an anonymous referee for constructive comments. NKP acknowledges current support from NASA’s SDO/AIA (NNG04EA00C) and previous support from NPP at the  NASA/MSFC, administered by USRA under contract with NASA.  A.C.S and R.L.M acknowledge the support from the NASA HGI program. S.K.T. gratefully acknowledges support by NASA contracts NNG09FA40C (IRIS), and NNM07AA01C (Hinode). We acknowledge the use of Hi-C 2.1, \iris, and \sdo\ data. AIA is an instrument onboard the Solar Dynamics Observatory, a mission for NASA’s Living With a Star program. We acknowledge the Hi-C 2.1 instrument team for making the second re-flight data available under NASA Heliophysics Technology and Instrument Development for Science (HTIDS) Low Cost Access to Space (LCAS) program).  MSFC/NASA led the mission with partners including the SAO, the UCLan, and LSMAL. Hi-C 2.1 was launched out of the White Sands Missile Range on 2018 May 29. \iris\ is a NASA small explorer mission developed and operated by LMSAL with mission operations executed at NASA Ames Research center and major contributions to downlink communications funded by ESA and the Norwegian Space Centre. 

%We acknowledge the High-resolution Coronal Imager (Hi-C 2.1) instrument team for making the second re-flight data available under NASA grant. MSFC/NASA led the mission with partners including the Smithsonian Astrophysical Observatory and the University of Central Lancashire. Hi-C 2.1 was launched out of the White Sands Missile Range on 2018 May 29.
% We thank the referee for constructive comments. B.D.P acknowledges support from NASA grants NNX16AG90G, and NNG09FA40C (IRIS).

%\bibliographystyle{aasjournal}
%\bibliography{sun}

\end{document}